\documentclass[nolinenumbers,twocolumn,trackchanges,times]{aastex631}
\usepackage{CJK}
\usepackage{xspace} 
\usepackage{amsmath}
\usepackage{url}
\usepackage{natbib}
\usepackage{tabularx}
\usepackage{color}
\usepackage{graphicx}
\usepackage{epstopdf}
\usepackage{gensymb}
\usepackage{hyperref}
\usepackage{longtable}
\usepackage{subfigure}
\usepackage[T1]{fontenc}
\DeclareUnicodeCharacter{2212}{-}

\newcommand*{\teff}{$T_{\rm eff}$\xspace}
\newcommand*{\logg}{$\log~g$\xspace}

\newcommand*{\kms}{km s$^{-1}$\xspace}
\newcommand*{\zmax}{$Z_{\rm max}$\xspace}

\newcommand*{\rmax}{$r_{\rm apo}$\xspace}
\newcommand*{\rmin}{$r_{\rm peri}$\xspace}

\newcommand*{\rperi}{$r_{\rm peri}$\xspace}

\newcommand*{\vthe}{$V_{\rm \theta}$\xspace}
\newcommand*{\vphi}{$V_{\rm \phi}$\xspace}

\newcommand*{\rsun}{$R_\odot$\xspace}

\xspace

\newcommand*{\gaia}{$\it Gaia$\xspace}

\newcommand*{\stackel}{St$\ddot{a}$ckel\xspace}

\begin{document}

\shorttitle{Identifying Substructures}
\shortauthors{Kim et al.}

\begin{CJK}{UTF8}{mj}
\title{A Novel Approach to Identifying Substructures Through Analysis of Metallicity Distribution Functions}

\author[0000-0002-6411-5857]{Young Kwang Kim}
\affiliation{Department of Astronomy and Space Science, Chungnam National University, Daejeon 34134, South Korea}

\author[0000-0001-5297-4518]{Young Sun Lee (이영선)}
\affiliation{Department of Astronomy and Space Science, Chungnam National University, Daejeon 34134, South Korea; youngsun@cnu.ac.kr}

\author[0000-0003-4573-6233]{Timothy C. Beers}
\affiliation{Department of Physics and Astronomy and Joint Institute for Nuclear Astrophysics -- Center for the Evolution of the Elements (JINA-CEE), University of Notre Dame, Notre Dame, IN 46556, USA}

\begin{abstract}

We present a new method for identifying Galactic halo substructures accreted from dwarf galaxies by combining metallicity distribution functions (MDFs) with orbital parameters. Using apogalactic distance--orbital phase space, we assume that the MDF peak of a substructure reflects its progenitor’s chemical signature. We test this approach with two Galactic potentials (\stackel and McMillan) and find consistent results. Our sample consists of retrograde halo stars with low orbital inclinations and intermediate eccentricities ($0.5<e\leq0.7$), drawn from SDSS and LAMOST spectroscopy combined with $Gaia$ DR3 astrometry. We identify four distinct low-inclination retrograde substructures (LRS 1, LRS 2, LRS 3, LRS 4) with MDF peaks at [Fe/H] = $-$1.5, $-$1.7, $-$1.9, and $-$2.1, respectively; LRS3 is newly discovered. Further analysis reveals an additional stream (LRS 2B) with [Fe/H] = $-$2.3 embedded within LRS 2; the remaining LRS 2 stars (LRS 2A) are associated with Sequoia. LRS 1 is likely linked to Thamnos 2 and Arjuna, and LRS 4 to I’itoi. Comparison with the ED-2 stream suggests LRS~2B is chemically distinct, but high-resolution spectroscopy is required to confirm whether they originate from separate progenitors. Our MDF-based approach demonstrates the utility of chemo-dynamical space for uncovering halo substructures, while highlighting caveats such as metallicity gradients and redshift evolution of the mass-metallicity relation, which may blur the mapping between MDF peaks and progenitors.
\end{abstract}


\section{Introduction} \label{sec:intro}

It is now well established that the Milky Way (MW) has assembled hierarchically through multiple mergers and accretion events of less massive galaxies (\citealt{white1991}). The stellar halo preserves the debris of these events as fossil records owing to the long dynamical timescales required for complete phase mixing (\citealt{Bland-Hawthorn2016,helmi2020}). Identifying and characterizing such substructures is therefore essential for reconstructing the MW’s formation and assembly history, requiring precise 6D phase-space data and chemical abundance measurements.

The advent of $Gaia$ Data Releases (\citealt{gaia2016,gaia2018,gaia2021,gaia2023}) has revolutionized this field, providing unprecedented astrometric data for millions of stars. When combined with large spectroscopic surveys such as legacy Sloan Digital Sky Survey (SDSS; \citealt{york2000}), the Sloan Extension for Galactic Understanding and Exploration (SEGUE; \citealt{yanny2009}; \citealt{rockosi2022}), the Large sky Area Multi-Object Fiber Spectroscopic Telescope (LAMOST; \citealt{cui2012}; \citealt{luo2015,luo2019}), the Apache Point Observatory Galactic Evolution Experiment (APOGEE; \citealt{majewski2017,abolfathi2018,abdurro'uf2022}), these data have enabled the discovery of major accretion remnants (e.g., $Gaia$-Sausage/Enceladus, GSE; \citealt{belokurov2018,helmi2018}) and numerous smaller substructures (\citealt{myeong2018,koppelman2019,naidu2020,dodd2023,kim2025}).

From a dynamical perspective, because the stellar total energy, angular momentum actions, and angular momentum perpendicular to the $Z$ component of angular momentum ($L_{\rm{Z}}$) are conserved under adiabatic changes of a Galactic gravitational potential (\citealt{binney2008,binney2012}), the merger remnants of a common origin should be clustered in integrals-of-motion (IoM) space (see \citealt{helmi1999,gomez2010,helmi2020}, and references therein). Based on this, numerous studies have been carried out to identify substructures or dynamically tagged groups accreted from dwarf galaxies in the local halo by applying clustering algorithms in the IoM space (e.g., \citealt{myeong2018,koppelman2019,borsato2020,gudin2021,lovdal2022,shank2022a,shank2022b,dodd2023,ou2023,cabrera garcia2024,zhang2024,kim2025}). However, simulations that include dynamical friction and evolving Galactic potentials show that IoM clumps can overlap: stars from a single progenitor may appear in multiple clumps, and debris from distinct progenitors (or heated in-situ stars) can occupy the same region (\citealt{jean-baptiste2017,khoperskov2023a,khoperskov2023b}).

Chemical abundances provide an important complementary dimension. High-resolution studies (e.g., \citealt{nissen2010,hayes2018}) have revealed that halo stars bifurcate into high- and low-[Mg/Fe] populations at [Fe/H] $\sim -1.5$, corresponding to in-situ and accreted origins, respectively. More recently, \citet{belokurov2022} and \citet{rix2022} have linked [Al/Fe] and metallicity trends to the early proto-disk (“Aurora”) population.

In this work, we focus on the metallicity distribution function (MDF) as a tracer of progenitor identity. Dwarf galaxies follow a mass-metallicity relation (\citealt{kirby2013}), so accreted systems of different masses are expected to have distinct MDF peaks. Moreover, \citet{jean-baptiste2017} showed that a single massive merger can produce multiple clumps in $E$--$L_{\rm Z}$ space; by grouping clumps with similar MDF peaks, our approach mitigates misclassification of debris from a single progenitor.

Nevertheless, several caveats must be considered. The mass-metallicity relation evolves with redshift (\citealt{deason2016,ma2016}), allowing dwarfs of different masses accreted at different times to share similar metallicities. Internal metallicity gradients (\citealt{koppelman2020,naidu2021,amarante2022,limberg2022}) can also create energy-dependent MDF shifts: metal-poor outer-halo stars are stripped earlier and occupy higher-energy orbits than the metal-rich inner core. Simulations further show that a single progenitor can produce multiple MDF peaks during successive stripping events (\citealt{amarante2022}). These effects mean that MDF peaks cannot be used in isolation to uniquely identify progenitors; robust interpretations must combine chemistry with orbital properties.

Guided by these considerations, we target retrograde halo stars with low orbital inclinations and intermediate eccentricities ($0.5<e\leq0.7$), a regime known to host several prominent substructures (e.g., Thamnos 2, Sequoia, Arjuna, I’itoi, ED-2; \citealt{koppelman2019,myeong2019,naidu2020,dodd2023,balbinot2024}). We search for substructures by analyzing their MDFs in apogalactic distance--orbital phase (OP) space, adopting apogalactic distance as a proxy for orbital energy. Because orbital parameters depend on the assumed Galactic potential, we perform the analysis under two widely used models.

Our goal is to test whether MDF peaks, when combined with dynamical information, provide a robust framework for identifying halo substructures and constraining their progenitors. The remainder of this paper is organized as follows: Section~\ref{sec:data} describes the data; Section~\ref{sec:orbit} details the orbital parameter calculations; Section~\ref{sec:sel} explains the selection of retrograde stars; Section~\ref{sec:results} presents the substructure identification and the discovery of a stellar stream; Section~\ref{sec:discussion} discusses the relationship to ED-2 and the effects of alternative distance scales; and Section~\ref{sec:summary} summarizes our conclusions.

\section{Data} \label{sec:data}

In this study, we constructed a combined SDSS and LAMOST DR6 (\citealt{zhao2012}; \citealt{wang2020})  (hereafter, SDSS/LAMOST) data set in the same manner as \citet{lee2023} (to which the interested reader is referred for details). The SDSS and LAMOST stellar sources are complementary to each other in the sense that the LAMOST stars are mostly brighter than $r_0$ $<$ 17 ($\sim 90 \%$ of our sample stars), whereas the SDSS stars cover the magnitude range $r_0$ = 14 -- 21. Thus, we can investigate different regions of the MW with stars over a wide range of magnitudes.

The SDSS data comprise stellar objects from the main legacy SDSS survey, as well as its sub-surveys, which are SEGUE, the Baryon Oscillation Spectroscopic Survey (BOSS; \citealt{dawson2013}), and the extended Baryon Oscillation Spectroscopic Survey (eBOSS; \citealt{blanton2017}). A recent version of the SEGUE Stellar Parameter Pipeline (SSPP; \citealt{allendeprieto2008}; \citealt{lee2008a,lee2008b,lee2011}; \citealt{smolinski2011}) delivers precise stellar atmospheric parameters from low-resolution ($R \sim 1800$) stellar spectra from these surveys. The typical uncertainties of the estimated parameters are 180 K for \teff, 0.24 dex for \logg, and 0.23 dex for [Fe/H]. The uncertainty in the derived [Mg/Fe] is generally $<$ 0.1 dex, as [Mg/Fe] covers a much smaller range than [Mg/H].

The SSPP is readily applicable to the LAMOST stellar spectra because of the similar wavelength coverage and spectral resolution, and can be used to estimate the stellar atmospheric parameters as well as chemical abundances such as [Mg/Fe], and [C/Fe] (see \citealt{lee2013,lee2015} for details). Following cross-matches with LAMOST DR6, the $g_0$, $r_0$, and $i_0$ magnitudes for the execution of the SSPP were taken from Pan-STARSS (\citealt{chambers2016}), SDSS, and the AAVSO Photometric All-Sky Survey (\citealt{Henden2018}).

From an analysis of the systematic differences in the stellar parameters and chemical abundances using about 44,000 common stars between SDSS and LAMOST DR6, we found mean offsets of 5 K, 0.04 dex, 0.1 dex, and 0.02 dex for \teff, \logg, [Fe/H], and [Mg/Fe], respectively. Since these are much smaller than the measured uncertainties, we did not correct the systematic offsets in the stellar parameters and chemical abundances.

In a similar way, we corrected systematic differences of +5.2 and +4.9 \kms in the SDSS and LAMOST DR6 radial velocities relative to $Gaia$ DR3, respectively, in order to match the $Gaia$ DR3 scale. In addition, we adjusted the systematic differences in distance moduli by comparing the photometric distances of the SDSS/LAMOST stars with parallax distances of the stars, using those with relative parallax errors smaller than 10\%, after correcting for the zero-point offset of $-$0.017 mas (\citealt{lindegren2021}). The corrected values from the SDSS and LAMOST are $-$0.083 and $-$0.011 mag for main-sequence (MS) and MS turn-off (MSTO) stars, and +0.043 and $-$0.148 mag for red giant stars, respectively. 

In addition, we removed a trend in the difference in the distance modulii, as a function of \logg\ and [Fe/H], using a linear correction function. The photometric distance of the SDSS stars was derived following the methods of \citet{beers2000,beers2012}, whereas that of the LAMOST stars was obtained from the value-added catalog of LAMOST DR7, derived by \citet{wang2016}. The final distance adopted for our program stars is the parallax distance if the relative parallax error is less than 20\%; otherwise, we apply the corrected photometric distance. Note that the parallax distance with a relative error of less than 20\%, after correcting for the zero-point offset of $-$0.017 mas, was adopted in various literature on substructure studies (\citealt{amarante2022,lovdal2022,malhan2022,ruiz-lara2022,dodd2023,balbinot2023,malhan2024}); hence it is used in this study for comparison with the results of those studies. The distance is suitable for a physical interpretation of the kinematic properties of stars (\citealt{malhan2024}) and provides reliable estimates of orbital parameters (\citealt{amarante2022}).

\section{Space-Velocity Components and Orbital Parameters} \label{sec:orbit}

To calculate space-velocity components in a spherical coordinate system for the SDSS/LAMOST stars, along with the proper motions from the \gaia\ DR3, we adopted $V_{\rm{LSR}}$ = 236 \kms\ (\citealt{kawata2019}) for the rotation of the local standard of rest (LSR), a Solar position of \rsun\ = 8.2 kpc (\citealt{Bland-Hawthorn2016}) in the disk plane from the Galactic center, and a vertical distance of $Z_{\odot} = 20.8$ pc from the mid-plane (\citealt{bennet2019}). The Solar peculiar motion with respect to the LSR was assumed to be ($U$, $V$, $W$)$_{\odot}$ = ($-$11.10, 12.24, 7.25) \kms\ (\citealt{schonrich2010}). In this notation, a star with \vphi\ $ > 0$ \kms\ has a prograde motion; retrograde rotation is indicated by \vphi\ $ < 0$ \kms. Stars with $V_{\rm{r}} > 0$ \kms\ move away from the Galactic center, and stars with \vthe\ $> 0$ \kms\ move toward the south Galactic pole. A positive $L_{\rm{Z}}$ indicates a prograde orbit. We computed orbital inclination of each star defined as ${\alpha} = \cos^{-1} (L_{\rm{Z}}/L)$ following \citet{kim2021}, where $L$ is the total angular momentum. A star with a range of $55^{\circ} < {\alpha} < 125^{\circ}$ moves with a high orbital inclination, while a star with ${\alpha} \leq 55^{\circ}$ has a prograde, low orbital inclination; a star with ${\alpha} \geq 125^{\circ}$ has a retrograde, low orbital inclination.

\begin{figure}[t]
\begin{center}
        \includegraphics[width=\columnwidth]{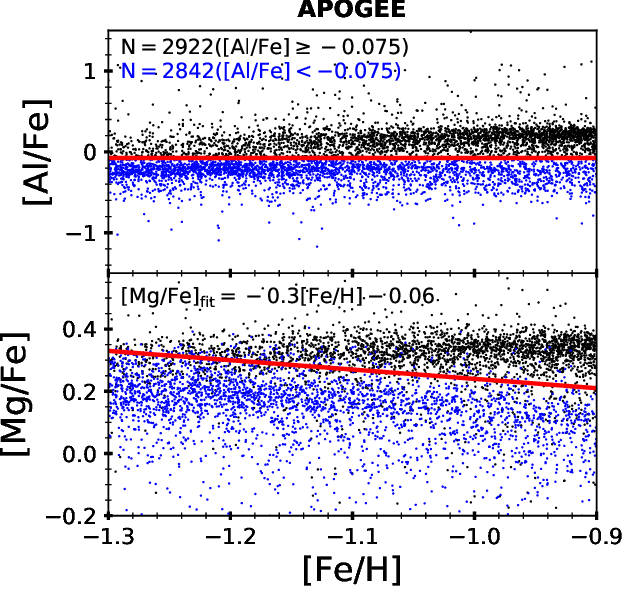}
	\caption{Distribution of APOGEE DR17 stars in the metallicity range $-$1.3 $<$ [Fe/H] $<$ $-$0.9 in the [Al/Fe]--[Fe/H] (top panel) and [Mg/Fe]--[Fe/H] (bottom panel) plane. These stars have uncertainties for both [Al/Fe] and [Mg/Fe] less than 0.05 dex and signal-to-noise ratio (S/N) greater than 100 in the $H$-band spectrum. The solid-red line is a reference line to separate the accreted stars from the in situ stars (see the text for the derivation of the line). In-situ stars with [Al/Fe] $\geq$ $-$0.075 are represented as black dots, while accreted ones with [Al/Fe] $<$ $-$0.075 are blue dots.}
	\label{figure1}
    \end{center}
\end{figure}

\begin{figure*}[t]
	\begin{center}
        \includegraphics[width=\textwidth]{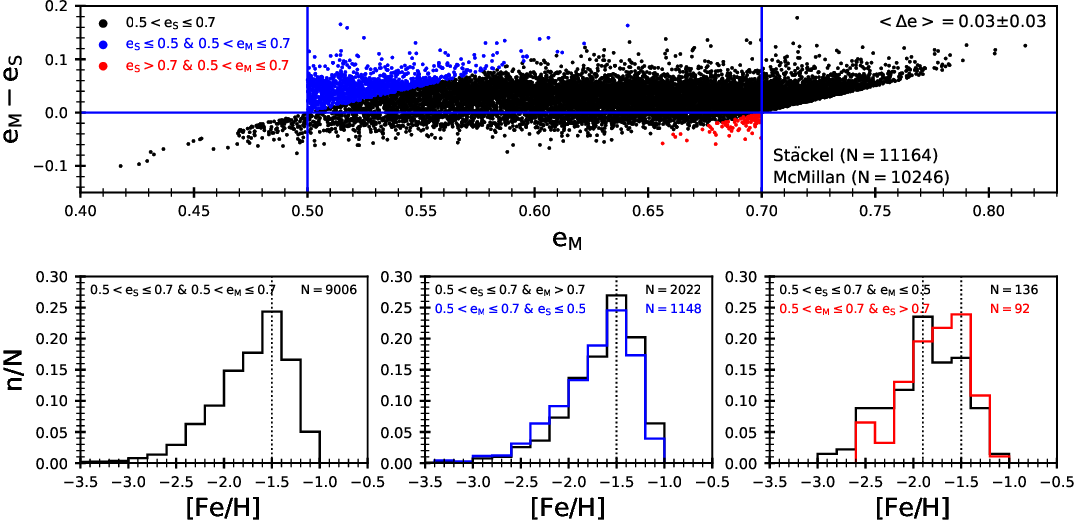}
		\caption{Top panel: Differences in the eccentricities of our sample stars between the \stackel\ ($e_{\rm{S}}$) and McMillan ($e_{\rm{M}}$) potentials, as a function of $e_{\rm{M}}$. The black dots are the stars in the eccentricity range $0.5 < e_{\rm{S}} \leq 0.7$ calculated from the  \stackel\ potential. The blue, black, and red dots between the two vertical blue lines represent the stars in the range $0.5 < e_{\rm{M}} \leq 0.7$ computed from the McMillan potential. Bottom panels: The lower-left panel is the MDF of the stars in the range $0.5 < e \leq 0.7$ in both potentials. The lower-middle panel shows the MDFs of the stars for ($e_{\rm{M}} - e_{\rm{S}}$) $\geq 0$. The black histogram is produced with the stars (black dots in the top-right corner of the top panel) in the ranges $0.5 < e_{\rm{S}} \leq 0.7$ and $e_{\rm{M}} > 0.7$, while the blue histogram with the stars (blue dots in the top panel) in the ranges $0.5 < e_{\rm{M}} \leq 0.7$ and $e_{\rm{S}} \leq 0.5$. Similarly, the lower-right panel is for ($e_{\rm{M}} - e_{\rm{S}}$) $< 0$. The black histogram is produced with the stars (black dots in the bottom-left corner of the top panel) in the ranges $0.5 < e_{\rm{S}} \leq 0.7$ and $e_{\rm{M}} \leq 0.5$, while the red one applies for stars (red dots in the top panel) in the ranges $0.5 < e_{\rm{M}} \leq 0.7$ and $e_{\rm{S}} > 0.7$.}	
		\label{figure2}
	\end{center}
\end{figure*}

We employed an analytic \stackel-type potential (hereafter, \stackel potential) with the tidal cutoff radius of 365 kpc (see \citealt{chiba2000,kim2019} for details) and a Galactic potential (hereafter, McMillan potential) from \citet{mcmillan2017}, to calculate the orbital parameters of the SDSS/LAMOST stars. These include the perigalactic distance (\rperi, the minimum distance of an orbit from the Galactic center), apogalactic distance (\rmax, the maximum distance of an orbit from the Galactic center), and stellar orbital eccentricity, $e$ = (\rmax $-$ \rmin) /(\rmax + \rmin), as well as \zmax\ (the maximum distance of an orbit above or below the Galactic plane). We used the software \texttt{AGAMA} (\citealt{vasiliev2019}) to compute the orbital parameters and angular momentum actions for the McMillan potential. Additionally, for both potentials, we computed the OP defined by ($r$ $-$ \rmin)/(\rmax $-$ \rmin) (\citealt{amorisco2015}), where $r$ is the distance of a star from the Galactic center. The OPs of 0 and 1 mean that a star is located at the perigalacticon (\rmin) or apogalacticon (\rmax), respectively. Uncertainties on the derived kinematic and orbital values were obtained from 100 Monte Carlo simulations with quoted uncertainties in the distance, radial velocity, and proper motions, assuming Gaussian error distributions.

\begin{figure*}[!t]
	\begin{center}
        \includegraphics[width=\textwidth]{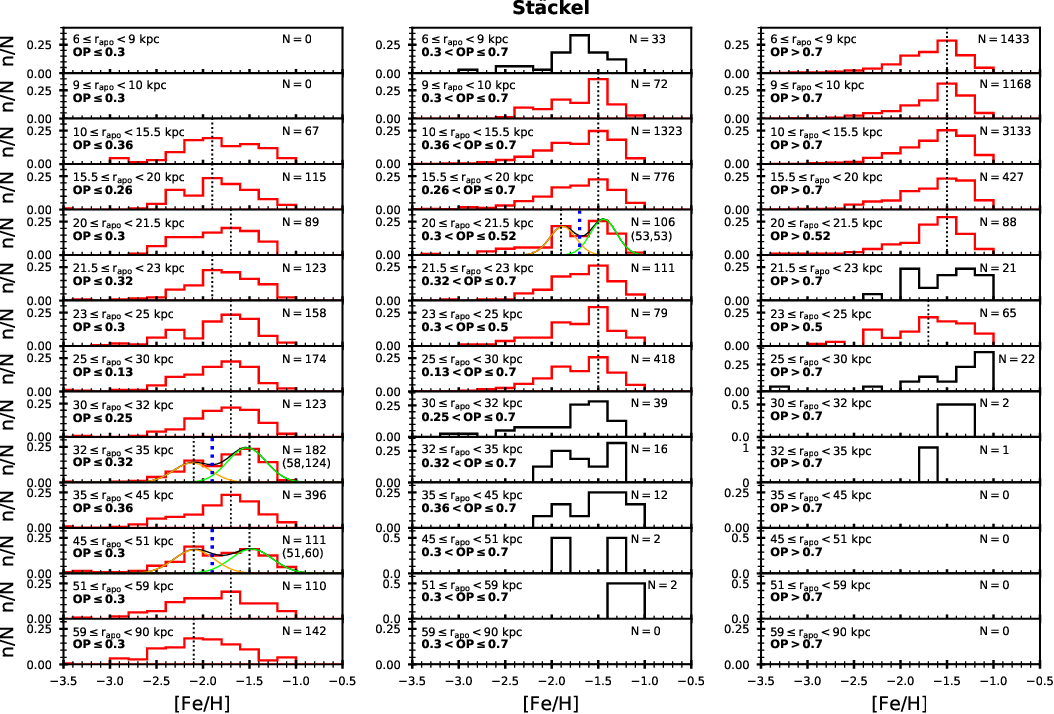}
	\caption{MDFs of the selected substructures (red histograms) in bins of \rmax and OP for the \stackel potential. Bins of \rmax increase from top to bottom; the OP bins increase from left to right in a given \rmax bin. The ranges of \rmax (black) and OP (bold black) are listed in the top-left corner of the respective panels. Because the shapes of MDFs are slightly different from Gaussian distributions, overlapping substructures are distinguished not by the division of the two Gaussian fits represented with the orange and green curves, but by the median [Fe/H] of the bin corresponding to the minimum between the two peaks. These [Fe/H] values are indicated by the vertical blue-dotted lines; the vertical black-dotted lines represent the MDF peaks of the selected substructures. See text for the adopted \rmax and OP ranges. Our analysis only considers the bins with $N$ $\geq$ 50.} 
		\label{figure3}
	\end{center}
\end{figure*}

\begin{figure*}[!t]
		\begin{center}
		\epsscale{1.1}
	\plotone{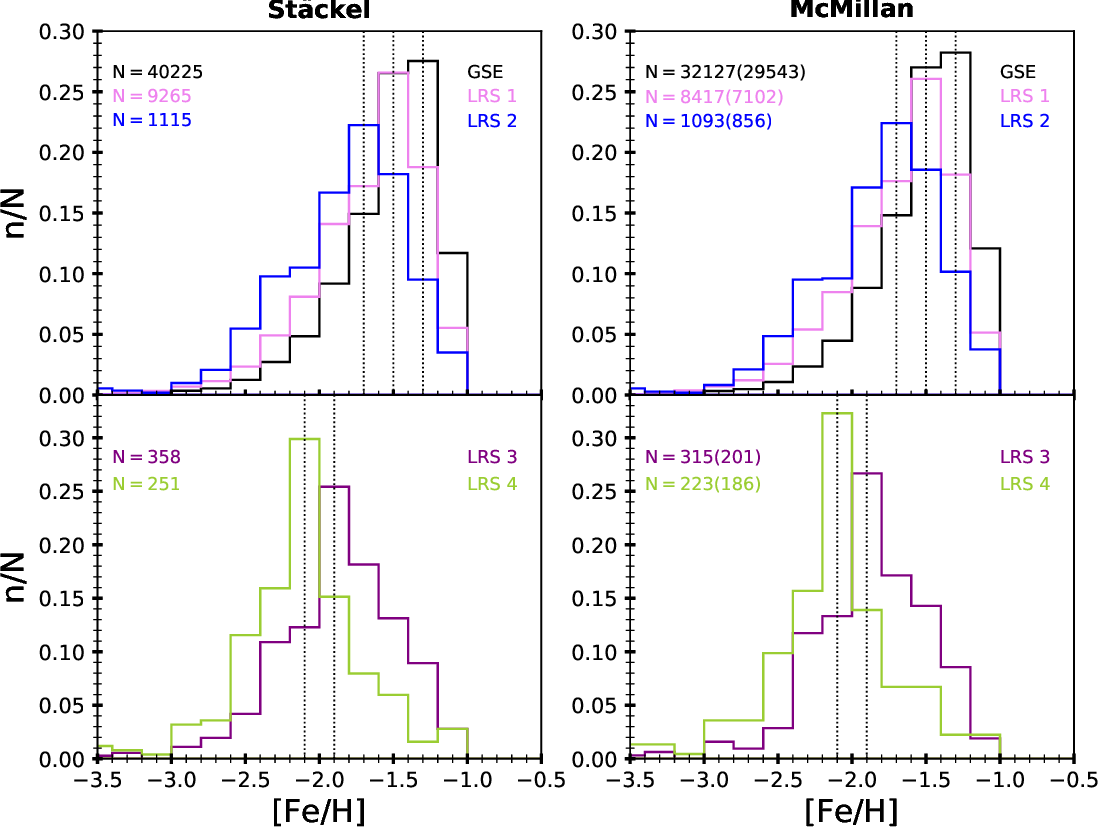}
	\caption{MDFs of each substructure selected in the \stackel\ potential (left panels) and McMillan potential (right panels). GSE is chosen by the selection method of \citet{horta2023} for the \stackel\ potential and that of \citet{feuillet2021} for the McMillan potential, with an additional cut of $e > 0.7$. The numbers in parentheses are the number of common stars in the \stackel and McMillan potentials for each substructure. The black-dotted vertical lines in the top-left and right panels for the \stackel and McMillan are the MDF peaks of LRS 1, LRS 2, and GSE. The black-dotted vertical lines in the bottom-left and right panels are the MDF peaks for LRS 3 and LRS 4 for the \stackel and McMillan potentials.}
	\label{figure4}
	\end{center}
\end{figure*}

\section{Elimination of In-situ Stars and Selection of Accreted Stars} \label{sec:sel}

Our sample consists of MS and MSTO stars from SDSS/LAMOST that meet the following criteria: average signal-to-noise ratio (S/N) in the wavelength range 4000 -- 8000 \AA\ $>$ 10, 0 $<$ $(g-r)_0$ $<$ 1.2, $4000 \leq T_{\rm{eff}} \leq 7000$ K, and $\log~g \geq 3.5$, with reliable estimates of stellar parameters and chemical abundances. Of these stars, we first excluded sample stars with renormalized unit weight error (RUWE) greater than 1.4, for high-quality astrometric data, and then removed those within $r < 5$ kpc from the Galactic center to avoid contamination by in-situ stars with low metallicity ([Fe/H] $<$ $-$1.3) such as Aurora (\citealt{belokurov2022,rix2022}) and Galactic bulge stars.

In addition, to eliminate in-situ stars born during proto-disk formation in the metallicity range $-$1.3 $<$ [Fe/H] $<$ $-$0.9 (\citealt{belokurov2022}), we derived a linear function of ${\rm{[Mg/Fe]}}_{\rm{fit}} = -0.3\cdot{\rm{[Fe/H]}} - 0.06$, using the [Al/Fe] and [Mg/Fe] values of the stars with S/N $>$ 100 and uncertainties less than 0.05 dex in APOGEE DR17. We regarded the SDSS/LAMOST stars that have [Mg/Fe] at a given [Fe/H] larger than ${\rm{[Mg/Fe]}}_{\rm{fit}}$ from the derived function as in-situ stars, and excluded them in our analysis.

The linear selection function was obtained as follows. We divided the metallicity interval of $-$1.3 $<$ [Fe/H] $<$ $-$0.9 into two ranges, $-$1.3 $<$ [Fe/H] $<$ $-$1.1 and $-$1.1 $<$ [Fe/H] $<$ $-$0.9. Then, following the similar assumption that the in-situ stars have [Al/Fe] $\geq$ $-$0.75 \citep{belokurov2022}, we determined upper bounds of +0.30 and +0.24 of [Mg/Fe] for the two regions, respectively. Each region includes 94\% of the accreted stars with [Al/Fe] $<$ $-$0.075. We then derived a linear function connecting the two points ($-$1.2, +0.30) and ($-$1.0, +0.24), which are the upper bounds and the median [Fe/H] values in each range. Figure~\ref{figure1} illustrates the results of this procedure. The figure shows the distributions of [Al/Fe] vs. [Fe/H] and [Mg/Fe] vs. [Fe/H] for the APOGEE DR17 stars in the top and bottom panels, respectively, where in-situ stars with [Al/Fe] $\geq$ $-$0.075 are represented in black and accreted stars with [Al/Fe] $<$ $-$0.075 in blue. The derived function is marked as red lines in each panel.

We found mean offsets of $-$0.047 and $-$0.016 dex for [Fe/H] and [Mg/Fe], respectively, between the SDSS/LAMOST data and those in APOGEE DR17, using 96,573 common stars. Because these are much smaller than the measured uncertainties, we applied the derived selection function to our data without correcting these offsets. 

Lastly, for each potential model, we gathered a final sample of retrograde stars with low orbital inclinations in the eccentricity range $0.5 < e \leq 0.7$ following the orbital-inclination classification of \citet{kim2021}. We excluded stars with [Fe/H] $\geq$ $-$1.0, where disk-system stars dominate, although accreted stars may still exist in this abundance regime. The number of stars finally selected is 11,164 for the \stackel and 10,246 for the McMillan potentials, respectively.

The top panel of Figure~\ref{figure2} shows the distribution of the difference in the calculated eccentricities of our final sample between the \stackel\ ($e_{\rm{S}}$) and McMillan ($e_{\rm{M}}$) potentials, as a function of $e_{\rm{M}}$. The black dots represent the eccentricity range of $0.5 < e \leq 0.7$ computed from the \stackel potential, while the blue, black, and red dots between the two vertical blue lines are in the range $0.5 < e \leq 0.7$ from the McMillan potential.

Figure~\ref{figure2} indicates that the eccentricities in the McMillan potential are systematically larger than those in the \stackel\ potential; the mean difference between the two potentials is 0.03. The bottom left panel of Figure~\ref{figure2} represents the MDF of common stars in both potentials. The middle panel shows the MDFs of the stars with ($e_{\rm{M}} - e_{\rm{S}}$) $\geq 0$, whereas the right panel corresponds to stars with ($e_{\rm{M}} - e_{\rm{S}}$) $< 0$. The conditions for selecting stars to construct the histograms are listed at the top of each panel. The bottom-middle panel suggests that there may exist just one substructure with an MDF peak at [Fe/H] = $-$1.5 in the range of $e_{\rm{S}} \leq 0.5$ (blue) for the \stackel\ potential and of $e_{\rm{M}} > 0.7$ (black) for the McMillan potential.

\begin{table*}[!t]
\caption{Criteria for Identifying Each Substructure in the \stackel\ and McMillan Potentials}
\label{table1}
\scriptsize
\begin{center}
\begin{tabular}{l|l|r|l|r}
\hline
\hline
         &~~~~~~~~~~~~~~~~~~~~~~~~~~~~~~~~~~~~~~\stackel\      &    $N~~$                &~~~~~~~~~~~~~~~~~~~~~~~~~~~~~~~~~~~~~~McMillan       &    $N~~$                                                                                  
\\
\hline

         & $~~6.0 \leq r_{\rm{apo}} < ~~9.0$ kpc \& OP $>$ 0.7       &   1,433                                      & $~~6.0 \leq r_{\rm{apo}} < ~~8.0$ kpc \& OP $>$ 0.75   &     444                                         \\
         & $~~9.0 \leq r_{\rm{apo}} < 10.0$ kpc \& 0.3 $<$ OP $\leq$ 0.7   &  72                                  & $~~8.0 \leq r_{\rm{apo}} < ~~9.6$ kpc \& 0.3 $<$ OP $\leq$ 0.75     &  1,679                       \\
         & $~~9.0 \leq r_{\rm{apo}} < 10.0$ kpc \& OP $>$ 0.7       &    1,168                                        & $~~8.0 \leq r_{\rm{apo}} < ~~9.6$ kpc \& OP $>$ 0.75     &   101                                         \\
         & $10.0 \leq r_{\rm{apo}} < 15.5$ kpc \& 0.36 $<$ OP $\leq$ 0.7   &  1,323                                & $~~9.6 \leq r_{\rm{apo}} < 14.2$ kpc \& 0.42 $<$ OP $\leq$ 0.72   &  953                             \\
         & $10.0 \leq r_{\rm{apo}} < 15.5$ kpc \& OP $>$ 0.7      &   3,133                                             & $~~9.6 \leq r_{\rm{apo}} < 14.2$ kpc \& OP $>$ 0.72    &  2,924                                           \\
    & $15.5 \leq r_{\rm{apo}} < 20.0$ kpc \& 0.26 $<$ OP $\leq$ 0.7  &  776                                        & $14.2 \leq r_{\rm{apo}} < 18.4$ kpc \& 0.31 $<$ OP $\leq$ 0.72    &   728                               \\
    LRS 1     & $15.5 \leq r_{\rm{apo}} < 20.0$ kpc \& OP $>$ 0.7 &   427                                            & $14.2 \leq r_{\rm{apo}} < 18.4$ kpc \& OP $>$ 0.72    &   566                                                 \\
         & $20.0 \leq r_{\rm{apo}} < 21.5$ kpc \& 0.3 $<$ OP $\leq$ 0.52 \& [Fe/H] $\geq$ $-$1.7 & 53  & $18.4 \leq r_{\rm{apo}} < 19.6$ kpc \& 0.4 $<$ OP $\leq$ 0.8 \& [Fe/H] $\geq$ $-$1.7  &  64  \\
         & $20.0 \leq r_{\rm{apo}} < 21.5$ kpc \& OP $>$ 0.52     &   88                                                & $19.6 \leq r_{\rm{apo}} < 21.5$ kpc \& 0.32 $<$ OP $\leq$ 0.6  &   132                                   \\
         & $21.5 \leq r_{\rm{apo}} < 23.0$ kpc \& 0.32 $<$ OP $\leq$ 0.7   &  111                                 & $19.6 \leq r_{\rm{apo}} < 21.5$ kpc \& OP $>$ 0.6  &  80                                                       \\
         & $23.0 \leq r_{\rm{apo}} < 25.0$ kpc \& 0.3 $<$ OP $\leq$ 0.7   &  79                                    & $21.5 \leq r_{\rm{apo}} < 23.7$ kpc \& 0.3 $<$ OP $\leq$ 0.53   &  113                                   \\
         & $25.0 \leq r_{\rm{apo}} < 30.0$ kpc \& 0.13 $<$ OP $\leq$ 0.7  &   418                                 & $23.7 \leq r_{\rm{apo}} < 29.2$ kpc \& 0.13 $<$ OP $\leq$ 0.7    &  458                                  \\
         & $32.0 \leq r_{\rm{apo}} < 35.0$ kpc \& OP $\leq$ 0.32 \& [Fe/H] $\geq$ $-$1.9 &  124         & $31.5 \leq r_{\rm{apo}} < 34.0$ kpc \& OP $\leq$ 0.32 \& [Fe/H] $\geq$ $-$1.9    &   97         \\
         & $45.0 \leq r_{\rm{apo}} < 51.0$ kpc \& OP $\leq$ 0.3  \& [Fe/H] $\geq$ $-$1.9  &  60          &  $45.0 \leq r_{\rm{apo}} < 54.5$ kpc \& OP $\leq$ 0.3 \& [Fe/H] $\geq$ $-$1.9   &  78            \\
\hline
         & $20.0 \leq r_{\rm{apo}} < 21.5$ kpc \& OP $\leq$ 0.3   &  89                                              & $18.4 \leq r_{\rm{apo}} < 19.6$ kpc \& OP $\leq$ 0.4   &   94                                                 \\
         & $23.0 \leq r_{\rm{apo}} < 25.0$ kpc \& OP $\leq$ 0.3  &  158                                              & $21.5 \leq r_{\rm{apo}} < 23.7$ kpc \& OP $\leq$ 0.3    &   149                                              \\
         & $23.0 \leq r_{\rm{apo}} < 25.0$ kpc \& OP $>$ 0.5  &     65                                                 & $21.5 \leq r_{\rm{apo}} < 23.7$ kpc \& OP $>$ 0.53   &  75                                                     \\
      LRS 2    & $25.0 \leq r_{\rm{apo}} < 30.0$ kpc \& OP $\leq$ 0.13  &  174                                   & $23.7 \leq r_{\rm{apo}} < 29.2$ kpc \& OP $\leq$ 0.13    &   163                                            \\
         & $30.0 \leq r_{\rm{apo}} < 32.0$ kpc \& OP $\leq$ 0.25    &   123                                         & $29.2 \leq r_{\rm{apo}} < 31.5$ kpc \& OP $\leq$ 0.25   &  135                                              \\
         & $35.0 \leq r_{\rm{apo}} < 45.0$ kpc \& OP $\leq$ 0.36    &   396                                         & $34.5 \leq r_{\rm{apo}} < 45.0$ kpc \& OP $\leq$ 0.36   &   398                                             \\
         & $51.0 \leq r_{\rm{apo}} < 59.0$ kpc \& OP $\leq$ 0.3   &   110                                            & $54.5 \leq r_{\rm{apo}} < 62.0$ kpc \& OP $\leq$ 0.25   &   79                                              \\
\hline
 & $10.0 \leq r_{\rm{apo}} < 15.5$ kpc \& OP $\leq$ 0.36    &   67                                                   & $~~9.6 \leq r_{\rm{apo}} < 14.2$ kpc \& OP $\leq$ 0.42   &  55                                             \\
LRS 3         & $15.5 \leq r_{\rm{apo}} < 20.0$ kpc \& OP $\leq$ 0.26  &   115                                   & $14.2 \leq r_{\rm{apo}} < 18.4$ kpc \& OP $\leq$ 0.31  &  96                                                \\
         & $20.0 \leq r_{\rm{apo}} < 21.5$ kpc \& 0.3 $<$ OP $\leq$ 0.52 \& [Fe/H] $<$ $-$1.7  &  53    & $18.4 \leq r_{\rm{apo}} < 19.6$ kpc \& 0.4 $<$ OP $\leq$ 0.8 \& [Fe/H] $<$ $-$1.7  &  59        \\
         & $21.5 \leq r_{\rm{apo}} < 23.0$ kpc \& OP $\leq$ 0.32  & 123                                             & $19.6 \leq r_{\rm{apo}} < 21.5$ kpc \& OP $\leq$ 0.32    &   105                                            \\
\hline
         & $32.0 \leq r_{\rm{apo}} < 35.0$ kpc \& OP $\leq$ 0.32 \& [Fe/H] $<$ $-$1.9   &  58              & $31.5 \leq r_{\rm{apo}} < 34.0$ kpc \& OP $\leq$ 0.32 \& [Fe/H] $<$ $-$1.9    &  47               \\
 LRS 4        & $45.0 \leq r_{\rm{apo}} < 51.0$ kpc \& OP $\leq$ 0.3 \& [Fe/H] $<$ $-$1.9   & 51        & $45.0 \leq r_{\rm{apo}} < 54.5$ kpc \& OP $\leq$ 0.3 \& [Fe/H] $<$ $-$1.9    &  67                 \\
         & $59.0 \leq r_{\rm{apo}} < 90.0$ kpc \& OP $\leq$ 0.3   &  142                                             & $62.0 \leq r_{\rm{apo}} < 95.0$ kpc \& OP $\leq$ 0.25   &  109                                             \\
\hline

\end{tabular}
\tablecomments{We combined all stars in individual \rmax, OP, and [Fe/H] ranges to construct each substructure. $N$ is the number of stars satisfying individual selection conditions.}
\end{center}
\end{table*}

\begin{figure*}[!t]
	\begin{center}
    \includegraphics[width=\textwidth]{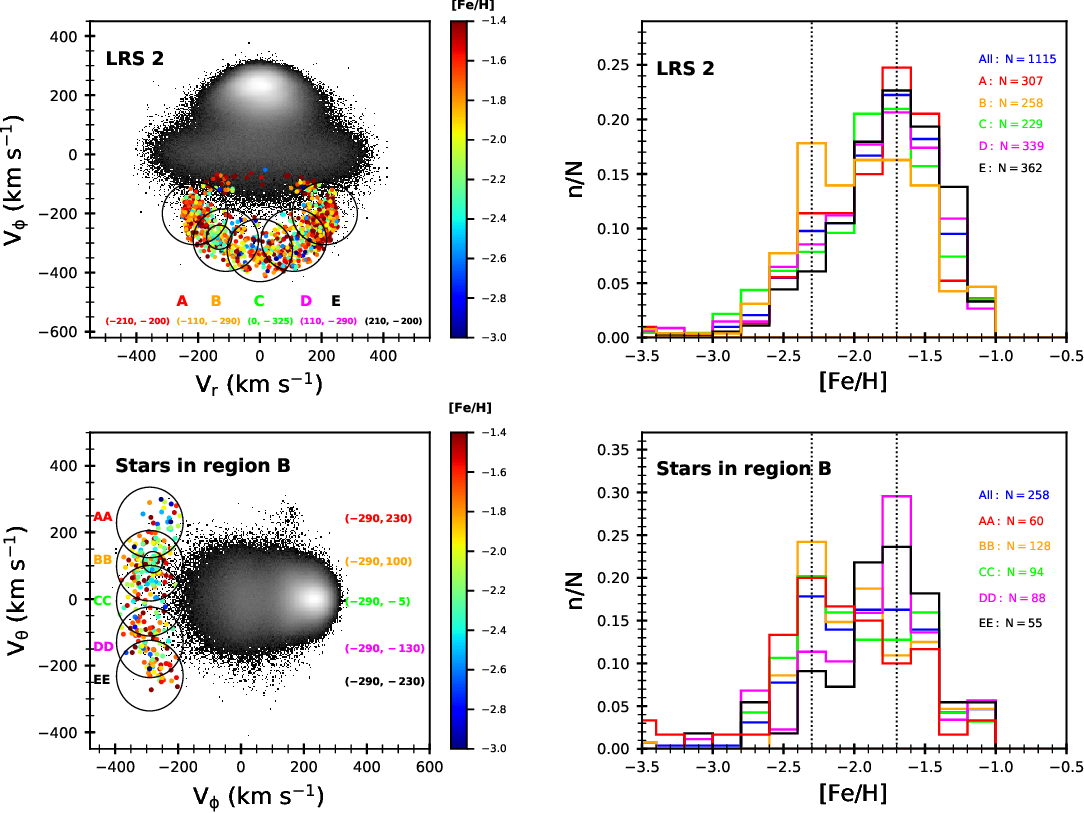}
	\caption{Top panels: Distribution of LRS 2 stars in the $V_{\phi}$ vs. $V_{\rm r}$ plane (left panel) and MDFs of stars in each bin of velocity space (right panel). The areas in velocity space in the left panel are represented as large circles with a radius of 106 \kms, with their centers (A, B, C, D, and E) listed at the bottom. The small circle marks the boundary of ED-2 from \citet{dodd2023}. Colored points indicate the metallicity of each star, coded by the color bar. Bottom panels: Similar to the top panels, but only for stars in the B region in the $V_{\theta}$ vs. $V_{\phi}$ plane. Note that these stars satisfy $\sqrt{[V_{\rm{r}}-(-110)]^2 + [V_{\phi}-(-290)]^2}$ $<$ 106 \kms\ in the top-left panel  (see text). The black-dotted vertical lines in the right panels are the MDF peaks for another stellar stream in LRS 2.}
	\label{figure5}
	\end{center}
	
\end{figure*}

\begin{figure}
	\begin{center}
        \includegraphics[width=\columnwidth]{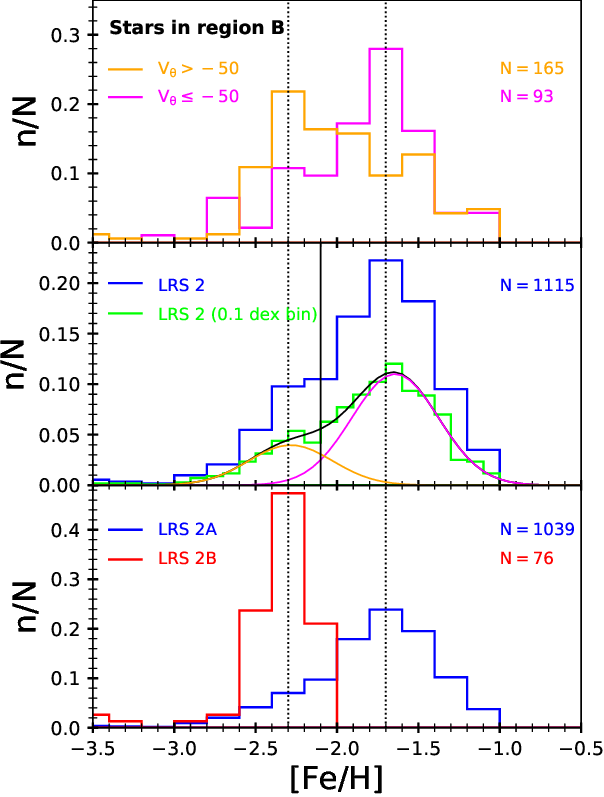}
		\caption{Top panel: MDFs of the LRS 2 stars in region B of Figure~\ref{figure5}, divided into $V_{\theta} > -50$ \kms (orange) and $V_{\theta} \leq -50$ \kms (magenta). Middle panel: MDFs of all LRS 2 stars with two bin sizes (blue with 0.2 dex bins and green with 0.1 dex bins). The orange and magenta curves denote best fits of the green histogram by two Gaussians with peaks at [Fe/H] = $-$2.28 and $-$1.64. The black curve represents the sum of the individual Gaussians. Bottom panel: MDFs of two proposed populations among the LRS 2 stars, dubbed LRS 2A (blue) and LRS 2B (red). LRS 2B stars are chosen by applying a [Fe/H] $<$ $-$2.1 cut (obtained by the Gaussian fits in the middle panel) to the stars of LRS 2 with $V_{\theta} > -50$ in the B region, whereas LRS 2A stars are the rest of LRS 2, excluding LRS 2B stars. The black-dotted vertical lines in each panel are the MDF peaks for a new stellar stream, LRS 2B and LRS 2.}
		\label{figure6}
	\end{center}
\end{figure}

\begin{figure}
	\begin{center}
        \includegraphics[width=\columnwidth]{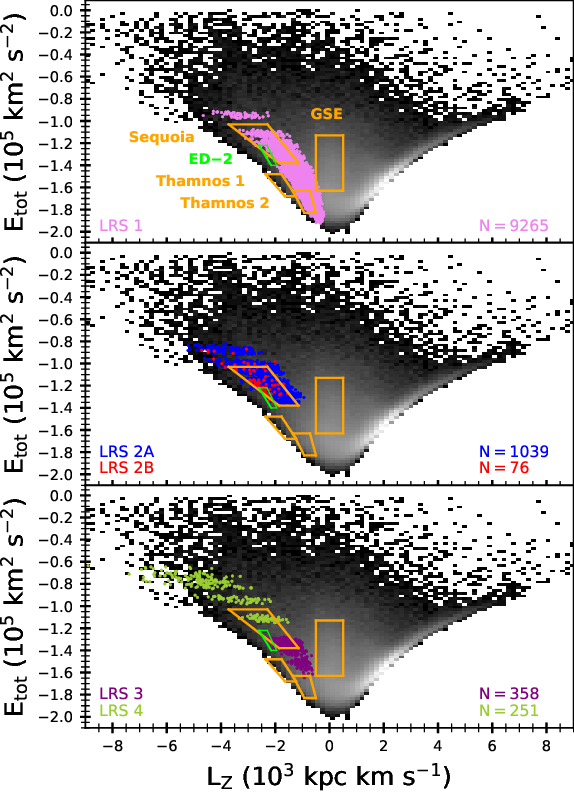}
		\caption{Distribution of stars in the $E_{\rm{tot}}$--$L_{\rm{Z}}$ plane for LRS 1 (top panel), LRS 2A and LRS 2B (middle panel), and LRS 3 and LRS 4 (bottom panel). The solid-orange boxes mark the locations of Sequoia, Thamnos 1, and Thamnos 2 used by \citet{koppelman2019}, and GSE by \citet{horta2023}, scaled to the \stackel\ potential used in this study. The solid-green box indicates the region of the ED-2 stars from \citet{dodd2023} and the ED-2 candidates from our sample.}
		\label{figure7}
	\end{center}
\end{figure}

\section{Results} \label{sec:results}
\subsection{Search for Substructures in the Apogalactic Distance and Orbital Phase Space (\rmax -- OP)} \label{subsec:Sea}

To identify potential substructures, we analyze the metallicity distribution functions (MDFs) of halo stars in the plane of apogalactic distance (\rmax) and orbital phase (OP). We begin by binning stars in \rmax starting from \rmax = 6 kpc, since only a small number of stars in our sample have \rmax $<$ 6 kpc. Within each \rmax bin, stars are divided into three OP intervals -- low (OP $\leq 0.3$), intermediate ($0.3 <$ OP $\leq 0.7$), and high (OP $> 0.7$) -- to examine their MDFs (Figure~\ref{figure3}). This framework allows us to track how metallicity peaks, which may indicate accreted substructures, evolve as a function of both orbital energy (approximated by \rmax) and phase.

The MDF in each bin is visually inspected to detect distinct metallicity peaks. In cases where MDF shapes are complex or peaks overlap, OP intervals are slightly adjusted to better isolate the features. If no peak is present in a bin with more than 50 stars, we expand the \rmax range incrementally and repeat the inspection. Once a peak is identified, we fix the OP binning and continue extending \rmax until either the MDF shape changes or a new peak emerges. Substructures are then assigned to the \rmax interval preceding the change, and the procedure is repeated for subsequent intervals. This iterative approach naturally produces uneven \rmax intervals across the panels of Figure~\ref{figure3}.

Note that we do not apply Gaussian mixture modeling (GMM) to identify peaks in the \rmax--OP space because the MDFs are often skewed and non-Gaussian, and GMM can yield spurious components when applied to such distributions. Instead, we locate substructures by visually identifying prominent MDF peaks and confirming their stability across adjacent bins. To minimize noise while preserving sensitivity to metallicity features, we adopt a bin size of 0.2 dex in [Fe/H], which corresponds to the typical measurement uncertainty.

Figure~\ref{figure3} shows the resulting MDFs for the adopted \stackel potential. The \rmax bins increase from top to bottom, and OP bins increase from left to right, with the bin ranges noted in the top-left corner of each panel. Black-dotted vertical lines indicate the metallicity peaks of the identified substructures, while blue-dotted lines denote boundaries separating overlapping components. Because the MDFs are not strictly Gaussian, these boundaries are defined by the minimum [Fe/H] value between adjacent peaks rather than the crossing point of Gaussian fits (orange and green curves shown for reference in the figure). We only consider bins with more than 50 stars to ensure statistical reliability in substructure detection.

Inspection of Figure~\ref{figure3} clearly reveals that the identified substructures have MDF peaks at [Fe/H] = $-$1.5, $-$1.7, $-$1.9, and $-$2.1.  The majority of the more metal-rich substructures are located near the apogalacticon (high OP) with small \rmax, while the stars located near the perigalacticon (low OP) exhibit broader MDFs.

We group together the MDFs exhibiting similar peaks in \rmax--OP bins in Figure \ref{figure3} to construct a unique substructure originating from a common progenitor, assuming that the substructures with similar MDF peaks share the same dwarf galaxy progenitor. We refer to the substructure with an MDF peak at [Fe/H] = $-$1.5 as low-inclination retrograde substructure (LRS) 1, at [Fe/H] = $-$1.7 as LRS 2, at [Fe/H] = $-$1.9 as LRS 3, and at [Fe/H] = $-$2.1 as LRS 4.

We apply the same procedure to the orbital parameters calculated from the McMillan potential. We compare the identified MDF in each \rmax bin with that from the \stackel\ potential and confirm a similar MDF shape in each \rmax and OP interval between the two potentials, albeit their parameter values and the number of stars differ slightly. Once again, the substructures with the same MDF peaks are grouped and named as in the \stackel potential. Table \ref{table1} summarizes the considered ranges of \rmax and OP for LRS 1, LRS 2, LRS 3, and LRS 4 for the \stackel\ and McMillan potentials.

Figure~\ref{figure4} shows the MDFs of LRS 1 (violet), LRS 2 (blue), LRS 3 (purple), and LRS 4 (yellow-green) from the \stackel (left panels) and McMillan (right panels) potentials.  The GSE substructure is selected by the method of \citet{horta2023} for the \stackel potential and by that of \citet{feuillet2021} for the McMillan potential, with an additional $e > 0.7$ cut. The numbers in parentheses are the number of stars in common for both potentials for each substructure. This figure again shows that the only difference between the two potential models is the number of stars in the selected substructures. Figures \ref{figure3} and \ref{figure4} confirm that we can draw a similar conclusion on the characteristics and interpretation of the substructures from the two adopted potential models; hence, we only consider the results of the \stackel\ potential to investigate and interpret the properties of each substructure in the following analysis.

\begin{figure*}[!t]
	\begin{center}
    \includegraphics[width=\textwidth]{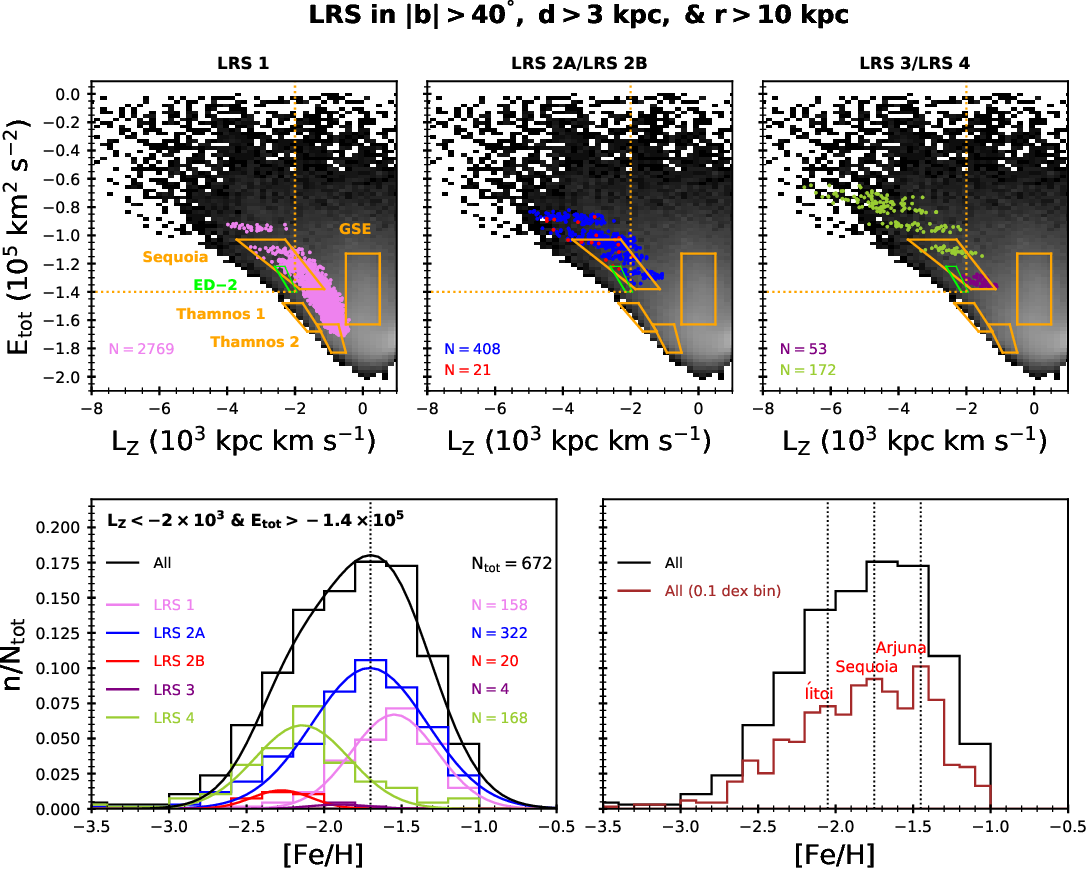}
    \caption{Top panels: Same as in Figure \ref{figure7}, but for stars with $|b|>40^{\circ}$, $d > 3~\rm{kpc}$ (distance from the Sun), which are the criteria used by \citet{naidu2020}, with an additional cut of $r > 10~\rm{kpc}$ (distance from the Galactic center) applied. Bottom panels: MDFs of individual substructures along with the Gaussian fits of those substructures and their sum (lower-left), and all substructure stars (right) in the region $L_{\rm{Z}} < -2 \times 10^3~{\rm{kpc}}~{\rm{km}}~{\rm{s}}^{-1}$ and $E_{\rm{tot}} > -1.4 \times 10^5~{\rm{km}}^2~{\rm{s}}^{-2}$ of stars in the top panels. The region of $L_{\rm{Z}} < -2 \times 10^3~{\rm{kpc}}~{\rm{km}}~{\rm{s}}^{-1}$ and $E_{\rm{tot}} > -1.4 \times 10^5~{\rm{km}}^2~{\rm{s}}^{-2}$ (dotted-orange box in the top panels) include the I'itoi, Sequoia, and Arjuna substructures with high energy and strong retrograde motion \citep{naidu2020}. In the lower-right panel, the MDF with 0.1 dex bin size (brown) distinctly shows at least three [Fe/H] peaks corresponding to Arjuna, Sequoia, and I'itoi, in order of metal-rich to metal-poor.}
	\label{figure8}
	\end{center}
\end{figure*}

\subsection{Discovery of a New Stellar Stream in the Velocity Space of LRS 2} \label{subsec:Dis}

In the previous subsection, we examined MDF peaks in various velocity spaces to substructures with possibly different origins rather than searching for streams with varying OPs in the velocity space of each group, as done by \citet{lovdal2022}. For an identified substructure, if there is a stellar stream with a different origin in a given velocity-space bin, we may expect more than one MDF peak. By examining the MDFs in various velocity regions for all selected substructures, we identify a new stellar stream associated with LRS 2.

To identify this new stellar stream, we first examine the distribution of LRS 2 stars in the 
$V_{\phi}$ -- $V_{\rm{r}}$ space, and the MDFs in five velocity-space regions (large circles), as shown in the top-left and right panels of Figure~\ref{figure5}, respectively. In the top-left panel, the large circles mark each velocity region determined by $\sqrt{(V_{\rm{r}}-V_{\rm{r0}})^2 + (V_{\phi}-V_{{\phi}0})^2}$ $<$ $R~(V_{\rm{r0}}, V_{{\phi}0})$, where ($V_{\rm{r0}}$, $V_{{\phi}0}$) denotes the region center given by A ($-$210, $-$200), B ($-$110, $-$290), C (0, $-$325), D (110, $-$290), and E (210, $-$200) \kms. $R~(V_{\rm{r0}}, V_{{\phi}0})$ is the radius of the circles (106 \kms). The colored points indicate the [Fe/H] values of the stars belonging to LRS 2, with colors scaled according to the accompanying color bar. The top-right panel clearly shows that region B's MDF (orange) has a distinct peak at [Fe/H] = $-$2.3, in addition to [Fe/H] = $-$1.7, while stars in other regions follow the MDF of LRS 2. This strongly suggests the presence of another stellar stream in the B region with a different origin from the rest of the LRS 2 stars.

To isolate the new stream in the B region for further investigation, we examine the distribution of the stars in the B region in the $V_{\phi}$--$V_{\theta}$ plane and the MDFs for five velocity space regions, as shown in the bottom left and right panels of Figure~\ref{figure5}, respectively. In the bottom-left panel, the large circles indicate the areas of each velocity area obtained by $ \sqrt{(V_{\phi}-V_{{\phi}0})^2 + (V_{\theta}-V_{{\theta}0})^2}$ $<$ $R~(V_{{\phi}0}, V_{{\theta}0})$, where ($V_{{\phi}0}$, $V_{{\theta}0}$) represents the region centers which are AA ($-$290, 230), BB ($-$290, 100), CC ($-$290, $-$5), DD ($-$290, $-$130), and EE ($-$290, $-$230) \kms. $R~(V_{{\phi}0}, V_{{\theta}0})$ is the circle radius (106 \kms). The metallicity of each star is indicated by the color bar. 

Inspection of the bottom-right panel of Figure~\ref{figure5} reveals that the MDFs of the AA, BB, and CC regions exhibit a peak at [Fe/H] = $-$2.3, whereas those of the DD and EE regions have a peak at [Fe/H] = $-$1.7, like LRS 2. A closer analysis suggests that the stars in the B region can be separated at $V_{\theta} = -50$ \kms\ located between CC and DD. To confirm this, we plot the MDF of the stars with $V_{\theta} > -50$ \kms\ (orange) in the B region, which correspond to the newly discovered stream and that of the stars with $V_{\theta} \leq -50$ \kms\ (magenta), as shown in the top panel of Figure \ref{figure6}. Furthermore, to examine how this stream is reflected in the MDF of LRS 2, we plot the MDF of LRS 2 with large (blue) and small (green) bin sizes in the middle panel of Figure \ref{figure6}. The MDF constructed with a 0.1 dex bin size exhibits two MDF peaks, and is well-reproduced by two Gaussians with peaks at [Fe/H] = $-$1.64 (magenta) and $-$2.28 (orange). The dividing point of the two Gaussians is at [Fe/H] = $-$2.1. Following this reasoning, we define the stars with [Fe/H] $<$ $-$2.1 and $V_{\theta} > -50$ in the B region as LRS 2B, a new stellar stream, while the rest of LRS 2 stars are referred to as LRS 2A. Their MDFs are displayed in the bottom panel of Figure~\ref{figure6}. We note that the small circles in the left panels of Figure~\ref{figure5} are the boundaries occupied by ED-2 stars from \citet{dodd2023}, which are within the velocity region of the new stellar stream, LRS 2B. Although the two substructures are likely to have different origins, as suggested by their minimal overlap in velocity space and by the metallicities reported for ED-2 in \citet{balbinot2024}, we provide a more detailed discussion of their possible relationship in Section \ref{sec:discussion}.

\begin{figure*}[!t]
	\begin{center}
        \includegraphics[width=\textwidth]{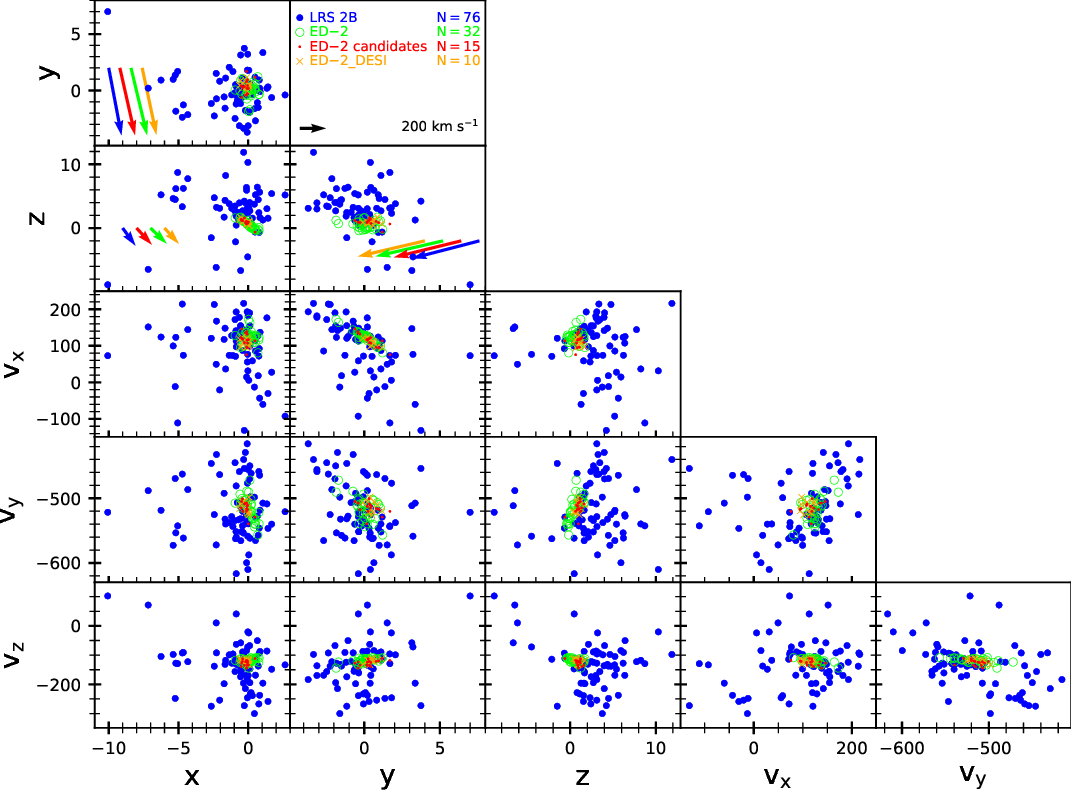}
		\caption{Spatial and velocity distributions of LRS 2B (blue), ED-2 (green) of \citet{dodd2023}, ED-2 candidates (red), and ED-2\_DESI sample (orange) from \citet{kim2025}, in heliocentric coordinates. In the top three panels, the blue, red, green, and orange arrows represent the mean projected velocity vectors for these substructures. This figure is displayed in the same coordinate system as that of \citet{balbinot2023} for comparison.}
		\label{figure9}
	\end{center}
\end{figure*}

\begin{figure}
	\begin{center}
        \includegraphics[width=\columnwidth]{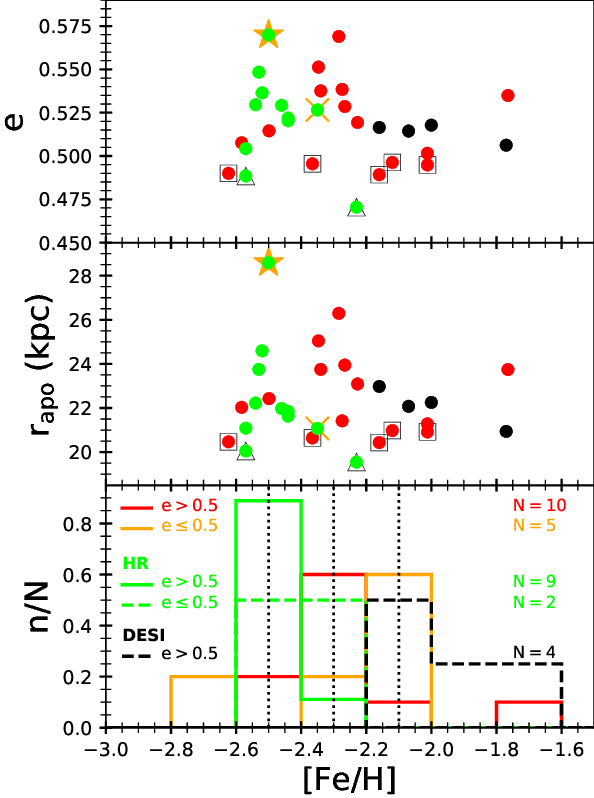}
		\caption{Top: Distribution of ED-2 stars in $e$ vs. [Fe/H]. Red dots represent the ED-2 candidates from our sample; the stars with $e \leq 0.5$ among them are plotted as black-open squares. The ED-2 members from high-resolution spectroscopic observations of \citet{balbinot2024} are shown with green dots; the stars with $e \leq 0.5$ are displayed as black-open triangles. The ED-2\_DESI sample with  $e > 0.5$ from \citet{kim2025} are marked with black dots. The single ED-2 member added by \citet{balbinot2023} is represented as an orange star. An orange cross indicates the $Gaia$ BH3 companion star from \citet{gaia2024}. Middle panel: Same as in the top panel, but in \rmax vs. [Fe/H]. Bottom panel: MDFs of our ED-2 candidates, the ED-2 members, and the ED-2\_DESI sample with $e > 0.5$. The MDFs of the ED-2 candidates with $e > 0.5$ from our sample are shown in solid-red lines; solid-orange lines are those with $e \leq 0.5$. Similarly, the MDFs of the ED-2 members with $e > 0.5$ from high-resolution spectroscopic observations of \citet{balbinot2024} are shown with solid-green lines; dashed-green lines are those with $e \leq 0.5$. The MDF of the ED-2\_DESI sample with $e > 0.5$ from \citet{kim2025} is shown with a dashed-black line. The black-dotted vertical lines are the MDF peaks for the ED-2 members with $e > 0.5$, our ED-2 candidates with $e > 0.5$, and our ED-2 candidates with $e \leq 0.5$.}
		\label{figure10}
	\end{center}
\end{figure}

\subsection{Association of our Identified Substructures with Thamnos 2, Arjuna, Sequoia, and I'itoi} \label{subsec:Rel}

Figure~\ref{figure7} shows the distributions of stars, from top to bottom, in LRS 1, LRS 2A and LRS 2B, and LRS 3 and LRS 4 in the $E_{\rm{tot}}$ vs. $L_{\rm{Z}}$ space. In this figure, the solid-orange boxes mark the areas of Sequoia, Thamnos 1, and Thamnos 2 defined by \citet{koppelman2019}, and GSE by \citet{horta2023}, after scaling to the \stackel\ potential. The small solid-green box indicates the region that includes the ED-2 stars from \citet{dodd2023} and the ED-2 candidates from our data (see Section~\ref{sec:discussion} for details). As seen in Figure \ref{figure7}, in the low-energy region where LRS 1 and LRS 3 are located, the box for Thamnos 2 is dominated by LRS 1. At the same time, about half of the 32 stars selected by \citet{naidu2020} as Thamnos 2 are in the range of $0.5 < e < 0.7$, and appear to be associated with LRS 3, which has an MDF peak at [Fe/H] = $-$1.9. In contrast, the high-energy region contains stars from all substructures. The region for Sequoia contains more stars from LRS 1 and LRS 2 than from LRS 3 and LRS 4. The ED-2 area is also populated with several stars from LRS 1, LRS 2A, and LRS 3, as well as some from LRS 2B.

To investigate the relationship of our substructures with Arjuna, Sequoia, and I'itoi identified by \citet{naidu2020} in the high-energy region, where the substructures overlap, as seen in Figure \ref{figure7}, we first select our substructures by application of their sample selection criteria: $|b|>40^{\circ}$, $d>3~\rm{kpc}$, and $r>10~\rm{kpc}$. We add the last condition, since most of their stars lie in that range. We plot our substructures selected by these criteria in the top panels of Figure~\ref{figure8}, similar to Figure~\ref{figure7}. Then, among the stars of these substructures, we choose those in the range $L_{\rm{Z}} < -2 \times 10^3~{\rm{kpc}}~{\rm{km}}~{\rm{s}}^{-1}$ and $E_{\rm{tot}} > -1.4 \times 10^5~{\rm{km}}^2~{\rm{s}}^{-2}$, adopted from \citet{naidu2020}. This range, indicated by the orange-dotted lines in the top panels of Figure~\ref{figure8} contains the high-energy, strongly retrograde stars of Arjuna, Sequoia, and I'itoi. In the bottom-left panel of Figure~\ref{figure8}, we plot the MDFs of the culled substructures as well as that for all of them, along with the Gaussian fits of each substructure and their sum. The MDF of all stars is presented in the right panel in two different bin sizes (0.1 dex for brown and 0.2 dex for black).

Inspection of the bottom-left panel of Figure~\ref{figure8} shows that, among the strongly retrograde substructures in the outer region of the Milky Way, three dominate: LRS 1, LRS 2A, and LRS 4, with MDF peaks at [Fe/H] = $-$1.5, $-$1.7, and $-$2.1, respectively. These substructures also exhibit three peaks in the MDF of all substructures (the brown histogram), produced with 0.1 dex bin size in the bottom-right panel of Figure~\ref{figure8}, corresponding to Arjuna, Sequoia, and I'itoi, respectively, even though there are a few stars with $e < 0.5$. Although Arjuna was identified as GSE by \citet{naidu2020}, the MDF of GSE in our sample is different from that of LRS 1, as shown in Figure~\ref{figure4}, and LRS 1 dominates within the Thamnos 2 low-energy region, as seen in the top panel of Figure~\ref{figure7}. Based on these results, Arjuna is considered to be a cluster of stars that escaped from LRS 1 with high energy during its accretion; hence, LRS 1 is more likely to be associated with Thamnos 2. LRS 1, LRS 2A, and LRS 4, which share similar energy and angular momentum with Thamnos 2, Sequoia, and I'itoi, respectively, have median values of [Fe/H] = $-1.59^{+0.25}_{-0.44}$, $-1.73^{+0.31}_{-0.45}$, and $-2.10^{+0.32}_{-0.40}$, respectively, where the dispersion is measured from the difference between the 16th and 84th percentiles of the MDF.

According to \citet{dodd2024}, despite the heterogeneity of samples used, the MDF of Thamnos 2 exhibits a peak near [Fe/H] = $-$1.50, with a pronounced metal-poor tail; in contrast, \citet{ruiz-lara2022} report a higher median value of [Fe/H] = $-$1.30 for this substructure. For Sequoia, the MDF peak is found at [Fe/H] = $-$1.60 in several studies (\citealt{myeong2019,matsuno2019,naidu2020}), while \citet{kim2025} report a median metallicity of $-$1.68 and \citet{ruiz-lara2022} report $-$1.50, slightly more metal-rich. Meanwhile, I'itoi shows an MDF peak at [Fe/H] = $-$2.10 in \citet{naidu2020}. In summary, our LRS 1, LRS 2A, and LRS 4 appear consistent with Thamnos 2, Sequoia, and I’itoi, respectively, while LRS 2B and LRS 3, which have distinct median metallicities of $-2.31^{+0.14}_{-0.19}$ and $-1.85^{+0.38}_{-0.40}$, respectively, represent newly identified substructures in this study.

We have identified substructures by assuming that substructures with the same MDF peak originate from the same satellite galaxy. However, the origin of the identified substructures may be interpreted differently depending on the following different viewpoints.

If the mass-metallicity relation evolves with redshift (\citealt{kirby2013,deason2016,ma2016}), galaxies at earlier epochs are more metal-poor at fixed stellar mass. Consequently, for two dwarf galaxies with the same mean [Fe/H], the earlier-accreted satellite must have been more massive than the later-accreted one. More-massive progenitors experience stronger dynamical friction and therefore sink to lower orbital energies more rapidly (e.g., \citealt{naidu2021,amarante2022}). In contrast, less-massive dwarf galaxies accreted later would remain at higher energies. As a result, stars from both progenitors could span a broad range of orbital energies while sharing a single MDF peak, as observed for LRS~1. This implies that the high-energy ($E_{\rm tot} > -1.4 \times 10^5~{\rm km}^2~{\rm s}^{-2}$) and low-energy components of LRS~1 may originate from distinct progenitors.

Considering that metallicity gradients in the progenitor of GSE can be imprinted in the orbital properties of its debris, metal-poor stars originating from the outskirts would have been stripped earlier and deposited onto higher-energy retrograde orbits. In contrast, more metal-rich stars from the inner regions, stripped later, would occupy lower-energy orbits with nearly zero net rotation due to radialization from dynamical friction (\citealt{koppelman2020,naidu2021,amarante2022,limberg2022}). Using the median [Fe/H] of $-1.47^{+0.22}_{-0.40}$ for GSE (selected following \citealt{horta2023}) and $-1.61^{+0.27}_{-0.40}$ for the high-energy stars of LRS~1, and adopting the structural parameters for the GSE progenitor from \citet{naidu2021}, we estimate a weak metallicity gradient of $-0.038^{+0.006}_{-0.007}~\rm dex~\rm kpc^{-1}$, consistent with \citet{limberg2022}. In contrast, using the most metal-poor substructure, LRS~2B, we infer a much stronger gradient of $-0.225^{+0.011}_{-0.008}~\rm dex~\rm kpc^{-1}$, similar to the low-feedback model of \citet{amarante2022}. The upper and lower values are measured from the difference between the 95\% confidence intervals of the 10,000-bootstrapped data.

The weak metallicity gradient is consistent with the idea that the high-energy stars of LRS~1 may share a common origin with GSE. LRS~4 and LRS~3, with an energy-dependent MDF change, may also have originated from the same satellite galaxy, while the low-energy stars of LRS~1 and LRS~2A may represent later stripping phases or a related progenitor component. Conversely, the strong gradient implied by LRS~2B suggests that this substructure, along with other high-energy debris (excluding the low-energy LRS~1 stars and LRS~3), could also originate from GSE, but represent an earlier stripping phase from its outermost regions. Low-energy LRS~1 and LRS~3, which have distinct MDFs and orbital properties, may instead trace separate progenitors.

\begin{figure*}[!t]
	\begin{center}
        \includegraphics[width=\textwidth]{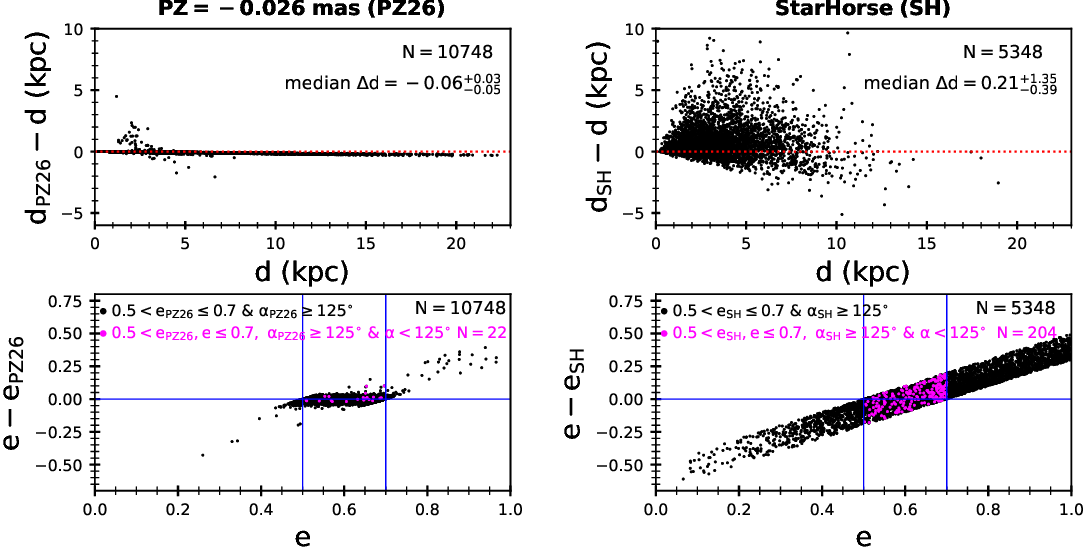}
		\caption{Left panels: The differences between $d_{\rm{PZ26}}$ and $d$, as a function of $d$, and the differences between the eccentricities obtained at those different distance scales, as a function of $e$, for the PZ26 sample. Right panels: Same as the left panels, but for the SH sample. Magenta dots are stars whose orbital eccentricities are within the selected range of eccentricities for both distance scales, but whose orbital inclinations obtained at this study's distance are outside the chosen range of orbital inclinations. All black dots between the two blue vertical lines, excluding the magenta dots, are stars in common with the kinematic-orbital selection space for both distance scales.}
		\label{figure11}
	\end{center}
\end{figure*}

\begin{figure*}[!t]
	\begin{center}
        \includegraphics[width=\textwidth]{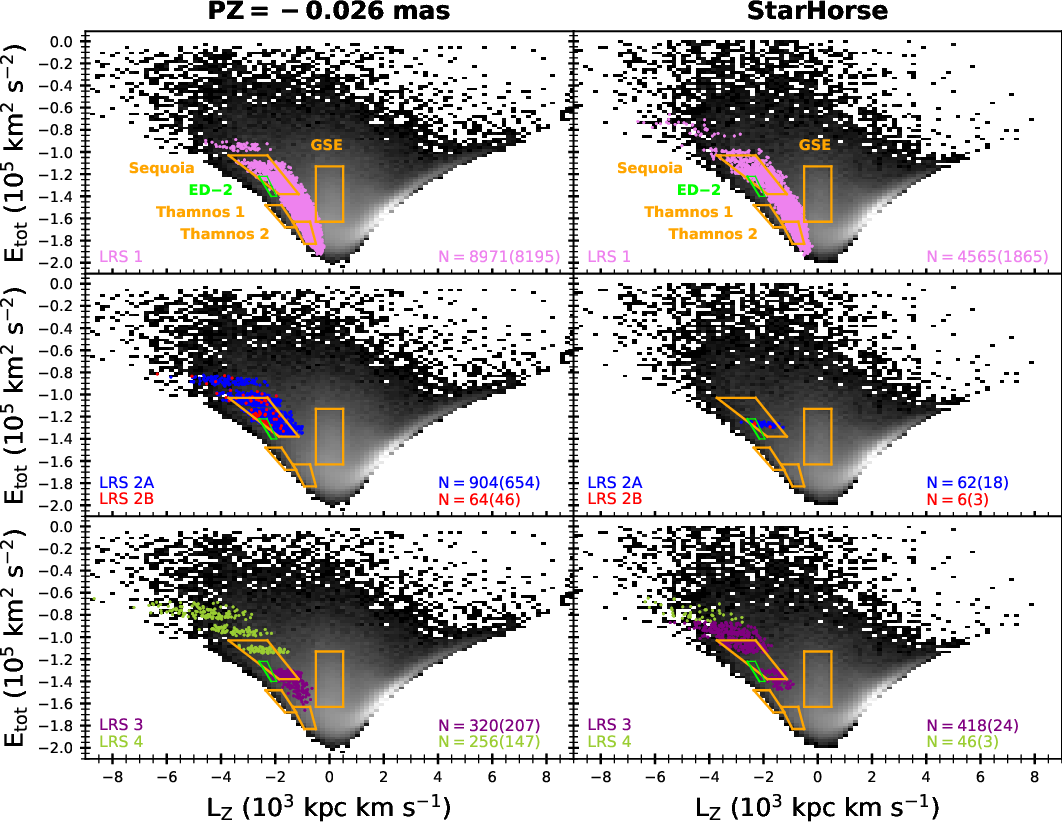}
		\caption{Same as in Figure \ref{figure7}, but for the PZ26 sample (left panels) and for the SH sample (right panels). For each of the left and right panels, the numbers in parentheses are the number of stars in common with the result from this study's distance for each substructure in the PZ26 sample and in the SH sample.}
		\label{figure12}
	\end{center}
\end{figure*}

\vskip 1 cm
\section{Discussion} \label{sec:discussion}
\subsection{Are LRS 2B and ED-2 Independent Structures?} \label{subsec:LRS 2B and ED-2}

Since LRS 2B and ED-2 overlap in velocity space, to confirm if the two substructures are related, we first compute the velocities and orbital parameters of the ED-2 stars identified by \citet{dodd2023} and the ED-2 stars (hereafter, ED-2 members) based on the high-resolution spectroscopic observations by \citet{balbinot2024}, using radial velocities, proper motions, and zero-point corrected parallax distances from $Gaia$ DR3, using the \stackel potential. Then, we extract from our SDSS/LAMOST data set the 15 ED-2 candidates that fall within the boundaries of the velocities ($V_{\rm{r}}-V_{\phi}, V_{\phi}-V_{\theta}$) and $L_{\rm{Z}}$ and angular momentum perpendicular to $ L_{\rm{Z}}$  ($L_{\perp} = \sqrt{L_{\rm{X}}^2 + L_{\rm{Y}}^2 } $, where $L_{\rm{X}}$ and $L_{\rm{Y}}$ are the $X$ and $Y$ components of angular momentum.) occupied by the ED-2 stars of \citet{dodd2023}. Recently, \citet{kim2025} have found a cluster E associated with ED-2 using the Dark Energy Spectroscopic Instrument (DESI) Milky Way survey data (\citealt{cooper2023}; \citealt{koposov2025}). Thus, we additionally cull 10 ED-2 stars (hereafter, the ED-2\_DESI sample) from cluster E in the same way as extracting our ED-2 candidates. There are no matching stars between our ED-2 candidates, the ED-2 members, and the ED-2\_DESI sample.

Figure~\ref{figure9} presents the distribution of positions and velocities in heliocentric coordinates for LRS~2B (blue dots) and three groups of ED-2 stars: the ED-2 stars from \citet{dodd2023} (small green circles), our ED-2 candidates (red dots), and the ED-2\_DESI sample from \citet{kim2025} (orange crosses). The mean projected velocity vectors for each group are shown as arrows in the corresponding colors in the top three panels. For consistency, this figure adopts the same coordinate system as \citet{balbinot2023}. As seen in the figure, the three ED-2 groups are tightly clustered in both position and velocity space, consistent with \citet{balbinot2023}, whereas LRS~2B stars exhibit a more extended distribution in both spaces, as reflected by their wider spread in the $E_{\rm tot}$--$L_{\rm Z}$ plane (Figure~\ref{figure7}). Nevertheless, the mean projected velocity vectors of the three ED-2 groups and LRS~2B are closely aligned, indicating that these populations are kinematically related and move with nearly the same speed and direction.

To examine the relationship between orbital parameters and metallicity for our ED-2 candidates (red dots), Figure~\ref{figure10} shows the distributions of eccentricity $e$ (top panel) and \rmax (middle panel) as a function of [Fe/H], together with the ED-2 members from \citet{balbinot2024} (green dots) and the ED-2\_DESI sample from \citet{kim2025} for $e>0.5$ (black dots). One orange star and one orange cross indicate, respectively, the additional ED-2 member identified by \citet{balbinot2023} and the $Gaia$ BH3 companion star from \citet{gaia2024}. Open black squares and triangles mark ED-2 candidates and ED-2 members with $e \leq 0.5$, respectively. The bottom panel presents MDFs for ED-2 candidates (solid-red/orange lines) and ED-2 members (solid- and dashed-green lines), separated by $e > 0.5$ or $e \leq 0.5$, along with the MDF of the ED-2\_DESI sample for $e > 0.5$ (dashed-black line).

From this figure, ED-2 candidates with $e \leq 0.5$ exhibit smaller \rmax and higher metallicities (MDF peak $\approx -2.1$) than those with $e > 0.5$ (MDF peak $\approx -2.3$), similar to LRS~2B. These differences suggest that the low- and high-eccentricity ED-2 candidates may have distinct origins. Moreover, while both the high-eccentricity ED-2 candidates and the ED-2 members share similar orbital properties with LRS~2B, their MDF peaks differ systematically by $\sim 0.2$ dex (bottom panel of Figure~\ref{figure10}), again hinting at possible differences in the progenitor or stripping history. The ED-2\_DESI sample with $e > 0.5$ also exhibit a somewhat higher metallicity, reinforcing this complexity.

For the two stars common to our ED-2 candidate list and the ED-2 members of \citet{balbinot2023}, which include metallicities from LAMOST DR3 (recalibrated by \citealt{li2018}) and LAMOST DR8, we find that our metallicities are offset by $-0.26$ dex relative to DR3 and by $+0.32$ dex relative to DR8. These offsets highlight the possibility of metallicity scale differences between surveys. A dedicated high-resolution spectroscopic follow-up of our ED-2 candidates and LRS~2B stars will be essential to establish whether ED-2 and LRS~2B are genuinely distinct substructures or different manifestations of the same progenitor.

More broadly, we caution that distinct MDF peaks do not necessarily imply distinct progenitors. Simulations of GSE-like mergers (e.g., \citealt{amarante2022}) show that a single massive dwarf can naturally produce multiple MDF peaks as stars are stripped during different orbital passages. This is particularly relevant for LRS~4, LRS~2A, and high-energy LRS~1, which occupy similar regions in $E_{\rm tot}$--$L_{\rm Z}$ space yet have distinct metallicities. Conversely, the relatively high metallicity of LRS~2A compared to the lower-energy LRS~3 is difficult to explain solely through metallicity gradients, favoring an interpretation involving multiple progenitors. These considerations emphasize the need to combine MDF analysis with additional chemical and kinematic diagnostics (e.g., $\alpha$-element abundances) to robustly identify progenitors.

\subsection{Impact of Different Distance Measures on the Substructure Search} \label{subsec:distance impact}

Different studies report varying zero-point corrections for $Gaia$ parallaxes (\citealt{riess2018,huang2021}). In addition, extensive work has shown that probabilistic approaches provide more reliable distances than simple parallax inversion, especially when relative parallax errors exceed 20\% (\citealt{bailer-jones2015}). Motivated by these findings, we test the robustness of our substructure identification against two alternative distance determinations: (i) distances corrected with the $-0.026$ mas zero-point from \citet{huang2021} (hereafter PZ26) and (ii) probabilistic spectrophotometric distances from the StarHorse catalog for LAMOST and SDSS (\citealt{queiroz2023}, hereafter SH). We compare these with the distances adopted in our main analysis.

For consistency, we define the same kinematic-orbital selection space as in our main analysis (retrograde stars with low orbital inclinations and $0.5 < e \leq 0.7$). Stars satisfying these criteria under the PZ26 distances are referred to as the PZ26 sample, and those selected under the SH distances as the SH sample. Distances ($d$), eccentricities ($e$), and orbital inclinations ($\alpha$) are subscripted accordingly (e.g., $d_{\rm PZ26}$ and $e_{\rm SH}$).

Figure~\ref{figure11} compares distances and eccentricities derived from the alternative scales with those from our main analysis. For the PZ26 sample, the median distance offset is negligible, $-0.06^{+0.03}_{-0.05}$ kpc, with 94\% of stars remaining within the same kinematic--orbital selection space. The slightly larger discrepancies for some stars arise because our main distances adopt photometric estimates (for large relative parallax errors), while the PZ26 distances use parallax inversion. In contrast, the SH distances exhibit a median offset of $0.21^{+1.35}_{-0.39}$ kpc, are systematically larger, and have substantial scatter; only 44\% of stars remain within the same selection space.

Figure~\ref{figure12} illustrates the $E_{\rm tot}$--$L_{\rm Z}$ distributions of substructures identified with the PZ26 and SH distances, analogous to Figure~\ref{figure7}. For the PZ26 sample, the substructure distributions closely resemble those in our main analysis: the fractions of overlapping members are 91\% for LRS~1, 73\% for LRS~2 (2A+2B), 65\% for LRS~3, and 58\% for LRS~4, and their median [Fe/H] values remain consistent. This indicates that small changes in the parallax zero-point have minimal impact on our results. Conversely, the SH sample shows substantial differences: LRS~2A is sparsely populated, LRS~3 dominates the high-energy region (and is nearly absent at low energy), and the overlap fractions with our main analysis drop to 41\%, 31\%, 6\%, and 7\% for LRS~1, LRS~2, LRS~3, and LRS~4, respectively. The metallicities also shift slightly, particularly for LRS~2B and LRS~3.

In summary, adopting the $-0.026$ mas parallax zero-point correction does not materially affect our substructure identifications, whereas using probabilistic StarHorse distances leads to significant membership changes and altered metallicity distributions. These findings highlight the sensitivity of halo substructure studies to distance determinations and the need for careful calibration of future distance scales.

\section{Summary and Conclusions} \label{sec:summary}

We selected MS and MSTO stars from SDSS and LAMOST spectroscopic data combined with \gaia DR3 astrometry, removing all stars with [Fe/H] $\geq -1.0$, where disk stars dominate, and excluding in-situ halo stars at [Fe/H] $< -1.0$. Our goal was to test a new approach for identifying halo substructures by analyzing MDFs in the \rmax--OP plane. To this end, we focused on retrograde stars with low orbital inclinations and intermediate eccentricities ($0.5 < e \leq 0.7$), a region known to host several prominent accreted substructures.

Within this sample, we searched for substructures occupying distinct regions of the \rmax--OP space and exhibiting unique MDF peaks, interpreted as signatures of different progenitor dwarf galaxies. Using both the \stackel\ and McMillan Galactic potentials, we identified four primary substructures (LRS~1, LRS~2, LRS~3, and LRS~4), with MDF peaks at [Fe/H] $\approx$ $-$1.5, $-$1.7, $-$1.9, and $-$2.1, respectively; the results are consistent across both potential models. Further MDF analysis revealed a new stellar stream within LRS~2, designated LRS~2B, with a peak at [Fe/H] $\approx$ $-$2.3, while the remaining LRS~2 stars are referred to as LRS~2A.

Of the five substructures identified, LRS~2B and LRS~3 are newly discovered in this study, whereas LRS~1, LRS~2A, and LRS~4 show properties consistent with Thamnos~2, Sequoia, and I’itoi, respectively. We interpret Arjuna as a high-energy component stripped from LRS~1 during its accretion. Because LRS~2B occupies the velocity space of the ED-2 stream, we compared our ED-2 candidates with the ED-2 members of \citet{balbinot2024} and found systematic abundance differences. A dedicated high-resolution spectroscopic follow-up of ED-2 candidates and LRS~2B stars will be required to confirm whether they represent distinct progenitors or related components.

Our MDF-based approach assumes that distinct metallicity peaks trace different progenitors; however, several caveats must be considered. First, metallicity gradients within dwarf galaxies can produce energy-dependent MDF variations, with metal-poor stars from the outer regions stripped earlier and occupying higher-energy orbits. Secondly, the mass-metallicity relation evolves with redshift (\citealt{deason2016,naidu2022,gonzalez-jara2025}), allowing progenitors of different masses accreted at different epochs to have similar mean metallicities. Thirdly, simulations show that a single progenitor can generate multiple MDF peaks due to successive stripping phases (\citealt{amarante2022}), as observed for GSE debris. These effects imply that MDF peaks alone cannot provide a one-to-one mapping between substructures and progenitors. We incorporate these considerations when interpreting LRS~4, LRS~2A, and high-energy LRS~1, whose overlapping orbital properties and complex MDFs may reflect different stripping phases of a single progenitor rather than distinct accretion events.

\begin{acknowledgments}

We thank an anonymous referee for a careful review of this paper, which has improved the clarity of its presentation. Y.K.K. acknowledges the support from the Basic Science Research Program through the NRF of Korea funded by the Ministry of Education (NRF-2021R1A6A3A01086446). Y.S.L. acknowledges support from the National Research Foundation (NRF) of Korea grant funded by the Ministry of Science and ICT (RS-2024-00333766). T.C.B. acknowledges partial support from grants PHY 14-30152; Physics Frontier Center/JINA Center for the Evolution of the Elements (JINA-CEE), and OISE-1927130: The International Research Network for Nuclear Astrophysics (IReNA), awarded by the US National Science Foundation.
 
Funding for the Sloan Digital Sky Survey IV has been provided by the Alfred P. Sloan Foundation, the U.S. Department of Energy Office of Science, and the Participating Institutions.

This work presents results from the European Space Agency (ESA) space mission Gaia. Gaia data are being processed by the Gaia Data Processing and Analysis Consortium (DPAC). Funding for the DPAC is provided by national institutions, in particular the institutions participating in the Gaia MultiLateral Agreement (MLA). The Gaia mission website is \url{https://www.cosmos.esa.int/gaia}. The Gaia archive website is \url{https://archives.esac.esa.int/gaia.}

The Guoshoujing Telescope (the Large Sky Area Multi-Object Fiber Spectroscopic Telescope, LAMOST) is a National Major Scientific Project which is built by the Chinese Academy of Sciences, funded by the National Development and Reform Commission, and operated and managed by the National Astronomical Observatories, Chinese Academy of Sciences.

\end{acknowledgments}

 \label{ref}
\end{CJK}
\end{document}